\documentclass[conference]{IEEEtran}

\pdfoutput=1
\usepackage{indentfirst}
\usepackage{amsmath}
\usepackage{mathtools}
\usepackage{amsfonts}
\usepackage{dsfont}
\usepackage{cite}
\usepackage{times}
\usepackage{graphicx}
\usepackage{subfigure}
\usepackage{bbm}
\usepackage{authblk}
\usepackage{enumerate}
\usepackage[noend]{algpseudocode}
\usepackage{xcolor}
\usepackage{booktabs}
\usepackage{caption}
\usepackage{authblk}
\usepackage{blindtext}
\usepackage{listings}
\usepackage{courier}
\usepackage[ruled, vlined, linesnumbered]{algorithm2e}

\title{On the Feasibility of Infrastructure Assistance to Autonomous UAV Systems}

\author[1]{Sabur Baidya\thanks{sbaidya@uci.edu}}
\author[1]{Marco Levorato\thanks{levorato@uci.edu}}
\affil[1]{Donald Bren School of Information and Computer Science, UC Irvine}
\affil[  ] {e-mail: {\textit {\{sbaidya, levorato\}@uci.edu}}}

\date{}

\begin{document}

\maketitle

\begin{abstract}
Infrastructure assistance has been proposed as a viable solution to improve the capabilities of commercial Unmanned Aerial Vehicles (UAV), especially toward fully autonomous operations. The airborne nature of these devices imposes constrains limiting the onboard available energy supply and computing power. The assistance of the surrounding communication and computing infrastructure can mitigate such limitations by extending the communication range and taking over the execution of compute-intense tasks. However, autonomous operations impose specific, and rather extreme in some cases, demands to the infrastructure. Focusing on flight assistance and task offloading to edge servers, this paper presents an in-depth evaluation of the ability of the communication infrastructure to support the necessary flow of information from the UAV to the infrastructure. The study is based on our recently proposed FlyNetSim, an open-source UAV-network simulator accurately modeling both UAV and network operations.
\end{abstract}

\section{Introduction}
\label{sec:introduction}
{Unmanned} Aerial Vehicles (UAV) have been the center of significant recent attention from the research community. The traditional focus on robotics aspects of these interesting airborne systems, such as flight dynamics, autonomy, navigation and group coordination~\cite{hoffmann2007quadrotor,jordan2006airstar,kim2004autonomous}, is being complemented by a growing body of work on innovative communication, networking and signal processing techniques supporting their operations~\cite{gupta2015survey,amorim2017radio,jawhar2017communication}.

Along the latter line of inquiry, a recent trend interconnects the UAVs to the Internet of Things (IoT) infrastructure to augment their capabilities. For instance, cellular networks can increase the communication range of the UAVs, and grant them access to the infrastructure. Edge or cloud servers can take over the execution of heavy-weight computing tasks, and coordinate the operations of multiple autonomous UAVs operating in the same area~\cite{callegaro2018optimal}.
Finally, through the infrastructure, UAVs can access data streams from external sensors, thus increasing their sensing accuracy and range~\cite{shaikh2018robust}. 

The objective of this paper is to assess the main challenges and trends in establishing an effective interconnection between the UAVs and the infrastructure. Intuitively, the degree of ``trust'' the UAV can afford to give to the infrastructure depends on a number of Quality of Service (QoS) metrics heavily influenced by network and environment parameters, such as the used protocol suite and technology, network load, building density, and UAV motion characteristics.
We explore a wide range of scenarios and identify parameter regions where remote control or assistance through the IoT infrastructure is feasible. We focus our investigation on two case-study applications: (\emph{i}) The UAVs transmit telemetry to a remote controller at the network edge to inform navigation control; (\emph{ii}) the UAVs offload computing tasks to an edge server.

In the former case, network impairments may impact delay or loss of telemetry packets, influencing the estimation error of UAVs' position or speed. An excessive estimation error may affect the ability of a remote controller to, for instance, determine disjoint navigation trajectories for multiple UAVs operating in the same area.

In the latter case, which refers to edge computing paradigm~\cite{bonomi2012fog}, the UAVs transfer data to be processed to an edge server connected to the communication infrastructure. Clearly, the tolerable delay and delay variations, as well as the frequency and distribution of trains of data packets to be delivered to the edge server depend on the level of support the UAV asks to the infrastructure.  For instance, the offloading of sporadic, although heavy-load, tasks at the application layer may not impose extremely stringent requirements to the network. Conversely, the offloading of the processing of signals used for fine-grain navigation control, may require a tight distribution of data delivery timing to result in an effective control.

In order to capture the characteristics and complexity of infrastructure-assisted UAV systems, we use the FlyNetSim simulator we presented in~\cite{baidya2018flynetsim,flynetsimgit}.
The FlyNetSim simulator integrates two open source simulators -- NS3~\cite{henderson2008network} and ArduPilot~\cite{team2016ardupilot, team2016sitl} 
-- to create a UAV-network simulator preserving the control stack and operations of real-world UAVs while enabling the detailed simulation of the surrounding network environment.

Our results emphasize how the performance of infrastructure-assistance are heavily influenced by the characteristics of the environment, such as the number of nodes active in the network, the distance, and the characteristics of the application and its corresponding traffic emission pattern. We conclude that effective infrastructure-assistance can be only achieved by enabling UAVs to use multiple technologies, and by implementing algorithms capable to switch from one to another during a mission.

The rest of the paper is organized as follows. Section~\ref{sec:background} summarizes current trends and approaches in UAVs communications amd infrastructure assistance to UAVs' operations. Section~\ref{sec:system} presents the system considered in this paper and illustrates the challenges associated with some case study scenarios of infrastructure-assistance. In Section~\ref{sec:simulator}, we provide a brief overview of FlyNetSim, a joint UAV-Network simulator we proposed in
~\cite{baidya2018flynetsim}. Section~\ref{sec:results} presents numerical results and a thorough discussion on the ability of the communication/computing infrastructure to support the autonomous operations of UAVs. Section~\ref{sec:conclusion} concludes the paper.

\begin{figure}[!t]
        \centering
        \includegraphics[width=1.0\linewidth]{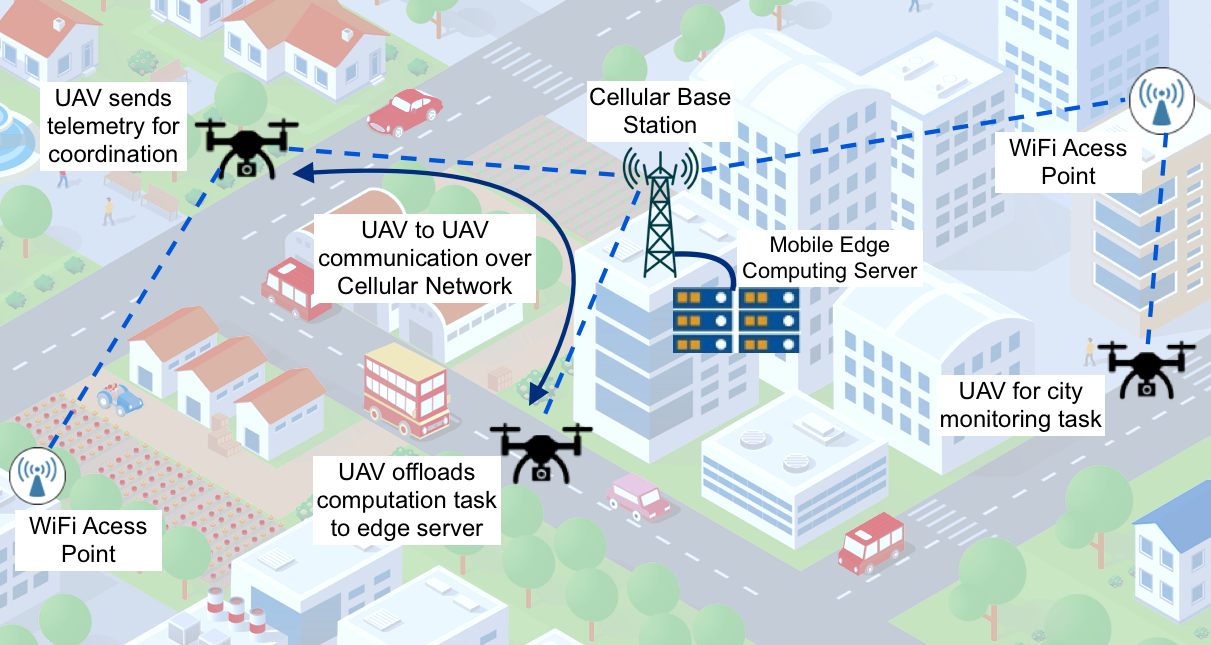}
\vspace{-4mm}
\caption{Scenario considered in this paper: UAVs connect to the communication/processing infrastructure to enhance their capabilities.} 
\label{fig:uav_city}
\vspace{-4mm}
\end{figure}
\section{Related Work}
\label{sec:background}



Autonomous UAVs necessitate to execute algorithms analyzing in real-time information rich signals, and extend their communication range to interconnect with remote flight coordinators or other UAVs. In the literature, many solutions and frameworks are available. We summarize the key approaches by dividing them into three categories: \emph{(a)} Infrastructure-assisted communications; \emph{(b)} Infrastructure-assisted computing; and \emph{(c)} Flying ad-hoc networks.


\subsection{Infrastructure-Assisted UAV Communications}

Clearly, urban area are at the same time the most challenging and infrastructure-rich environments. Thus, the UAVs have the opportunity to obtain help from available base stations and users, but also face many communication issues at different layers, including poor and/or unstable signal gain and contention from other active users.

Several contributions address the problem of interconnecting the UAVs with the infrastructure.
\cite{gu2015airborne} proposes to use properly aligned directional antennas to extend the range of UAVs communications over WiFi.  An experimental evaluation of UAV communications based on several WiFi standards was presented in~\cite{hayat2015experimental}. However, the aforementioned works do not consider UAV specific applications and the impact of UAV-specific characteristics on communication performance. In \cite{shaikh2018robust}, the authors presented a multi-hop communication strategy with dynamic make-before-break mechanism. The framework uses probes to evaluate available paths and select the one providing performance matching the needs of the UAV.

Cellular infrastructure-based UAV communications were proposed in~\cite{zeng2019cellular}, where aerial and ground UEs coexist in the same area. To increase the reliability of the cellular connection of the UAVs, an interference cancellation approach was used in~\cite{nguyen2018ensure}. A communication strategy over unlicenced bands was derived in~\cite{athukoralage2016regret}, where a regret function is defined to learn optimal duty cycle selection and support the coexistence of WiFi with other active technologies. Different from these contributions, herein we study the feasibility of using the infrastructure for relevant UAV based applications over available network technologies and different channel conditions.

\subsection{Infrastructure-Assisted Computing} 

To mitigate the limitations of on-board computation on the UAVs, the infrastructure can assist by taking over computation tasks necessary for autonomous UAV operations. By placing compute-capable devices at one -- wireless -- hop distance from mobile devices, edge computing offers low-latency services compared to traditional cloud computing.
Remarkably, offloading to edge servers not only has the potential for improving computing quality -- for instance by executing more complex algorithms -- and/or delay compared to on-board options, but also is a natural platform to support coordination across multiple UAVs.

Several contributions address challenges related to edge computing for UAVs. A framework to determine trajectories optimal both from the point of view of offloading and mission was proposed in~\cite{cao2018mobile}. The framework also accounts for constraints on the speed of the UAV. A hierarchical offloading approach was presented in~\cite{vega2015resilient} with real-world UAV-embedded setup for a computer vision based computing application. A framework for UAV-cloud computing for disaster rescue applications was proposed in~\cite{luo2015uav}. In this paper, we investigate the impact of the network and network parameters on task offloading.

\subsection{Flying Ad-Hoc Networks}
When the UAV operates beyond the range of ground resources, Flying Adhoc Networks (FANET)~\cite{bekmezci2013flying} are a suitable technologies to interconnect the UAVs and allow cooperation and coordination. Importantly, the low latency of direct UAV-to-UAV communications may improve the ability of a swarm to coordinate.

Note that due to UAV mobility, frequent reconfiguration might be needed when ad hoc communications are used for range extension or relaying of data streams.
Despite this issue, using intermediate relay nodes can improve the communication quality and range, especially as UAVs can form 3D topologies to avoid the impairments that characterize urban environments.

One of the main challenges in achieving effective communications is the distributed nature of flying ad-hoc networks. This may impair performance especially in dense deployments. To improve the performance of FANETs, ~\cite{xiao2016enabling} proposes to use millimeter-wave cellular communications with beamforming. However, the implementation of such model on real-world UAV is extremely challenging, especially due to the rapidly changing topology induced by moving UAVs.


\section{Infrastructure-Assisted UAV System}
\label{sec:system}

We consider an UAV operating in an urban environment, where multiple wireless access networks coexist. Specifically, we focus our attention on widely used communication technologies, that is, Wi-Fi and Long-Term Evolution (LTE) networks, which may support the intense transfer of data necessary to assist UAV operations. We do not assume that a channel is dedicated to the traffic generated by the UAV but, instead, include in the environment the activity of wireless nodes operating on the same channel resource. This will enable the study of the interactions between data streams induced by networking protocols at the different layers of the stack.

The UAV is assumed autonomous, meaning that it is not directly controlled by a human operator. However, some functionalities can be delegated to the infrastructure, that is, the UAV might depend on the help of other remote devices. In general, autonomy requires the acquisition and processing of signals from the surrounding environment to inform control. This transformation of the input signals to the output control has a rather broad meaning. The UAV can be simply programmed to follow a trajectory through a series of waypoints, in such case the input signal are GPS coordinates, and the UAV determines its motion to reach the next waypoint from its current position.  However, in many practical use-cases, an autonomous UAV may need to acquire and process complex signals, a task which may impose a significant burden to the battery-powered UAV in terms of energy expense. Furthermore, due to weight constraints UAVs often have weak on-board computing platforms, which may lead to a large time needed to execute compute-intense algorithms. Thus, offloading to compute-capable edge servers may lead to considerable energy consumption reduction as well as a faster capture to control pipeline for autonomy.

We provide in the following a short summary description of the class of infrastructure assistance problems we consider in this paper. The UAV produces a series of bursts of data at the application layers. We define, then, the series of instants $\{\tau_i\}_{i{=}1,2,3,\ldots}$, where $\tau_i$ corresponds to the emission time of the $i$--th burst. The bursts are characterized by a variable set $\mathit{B}_i$ that determines the amount of data generated in the $i$--th burst. We abstract the communication between the UAV and the access point -- Wi-Fi access point or LTE eNodeB in the considered environment -- through the transformation $\Phi$, which maps the emission time $\tau_i$ and burst size $B_i$ into the vectors ${\bf t}_i{=}[t_i(1), \ldots , t_i(N_i)]$ and $\boldsymbol{\omega}_i{=}[\omega_i(1), \ldots , \omega_i(N_i)]$. The vector ${\bf t}_i$ contains the delivery times of the $N_i$ network packets associated with the burst, the element $\omega_i(n)$ of $\boldsymbol{\omega}_i$ is equal to $1$ if the packet is delivered and $0$ otherwise. For convenience, we set to $\infty$ the delivery time of a failed packet. We define as $\Delta_i$ the difference between the delivery time of the packet of burst $i$ delivered the latest and the generation time $\tau_i$.

Intuitively, the transformation $\Phi$ is a function of a number of variables describing the complex state of the network. Clearly, analytically modeling the interactions between channel physical, access, link and transport layer protocols implemented by all the active nodes, as well as their packet arrival processes, is an impossible task, and we rely on detailed simulations to characterize the transformation $\Phi$ associated with specific communication strategies and environmental conditions.

\subsection{Network Environment}


In urban network infrastructure, the most commonly used wireless technologies are WiFi and LTE. WiFi is traditionally operated in unlicensed frequency bands, whereas LTE primarily operates in licensed spectrum -- although recent propositions extend its usage in the unlicensed spectrum as well. 
WiFi is a bidirectional communication technology specified in the IEEE 802.11 standard. In common infrastructure, WiFi stations (users) connect to an access point (AP) to communicate with other users or a remote server. The most used Medium Access Control (MAC) layer is Distributed Coordination Function (DCF), where the users contend for channel access using channel sensing and random backoff. Recent standards of WiFi support bitrate adaptation to improve the reliability of communications and Orthogonal Frequency Division Multiplexing (OFDM) modulation to increase capacity.

LTE is a cellular technology which defines an access network component called Evolved Universal Terrestrial Radio Access Network (E-UTRAN) and a core network component called Evolved Packet Core (EPC). The LTE access network has separate uplink and downlink radio bands, where resource allocation is controlled by the base station (eNodeB). The radio resources in LTE is allocated at the granularity of individual Physical Resource Block (PRB) which is the minimum unit that can be assigned to one user. Each PRB corresponds to one slot in the time domain and contains $12$ subcarriers of $15$ kHz each, thus occupying $180$~KHz in the frequency domain. The eNodeB scheduler allocates a subset of PRBs to each user for each subframe based on a scheduling policy. In Uplink, the PRBs allocated to a specific user should be contiguous, whereas in downlink the eNodeB can allocate non-contiguous PRBs to a user. Additionally, as the transmission power of the User Equipment (UE) is lower compared to that of the eNodeB, uplink is performed using Single Carrier FDMA (SC-FDMA) -- also known as DFT pre-coded OFDMA -- which provides a better peak-to-average power ratio. The instantaneous data transmission rate over the LTE physical channel is determined by the the transport block size (TBS) which is a function of the number of PRBs allocated to the UE and the Modulation and Coding Scheme (MCS) index. The transmitter receives the Channel Quality Index (CQI) information through a control channel and, based on the CQI level, it selects the appropriate MCS index.                                         

In the urban infrastructure, both WiFi and LTE networks are available to assist the UAV to enable the autonomous mission. However, the different characteristics of the protocol of the two technologies introduces different challenges in terms of protocol overhead, reliability, handling mobility, congestion and interference.

\subsection{Infrastructure Assistance}

Within this general environment, we consider the two specific infrastructure-assisted models described in the following. 

\vspace{2mm}
\noindent
{\bf Remote Navigation Assistance:} In a scenario where multiple UAVs share the same space, a remote unit can assist their navigation to avoid collisions, for instance by creating safe corridors for each UAV. To this aim, the UAV periodically transmits telemetry variables -- GPS coordinates, altitude, speed and battery level 
and other information if enabled on the UAV -- to the remote controller. The telemetry information are short messages usually contained within the maximum transmission unit (MTU) size of one network packet. The packets are usually sent at some update interval based on the polling request set by the GCS to fetch the telemetry information. Let's denote as ${\bf s}(t){=}(p(t),v(t),b(t))$ the state of the UAV at time $t$, where $p(t)$, $v(t)$ and $b(t)$ are position, speed, and battery level, respectively. In our model, the emission of telemetry variables corresponds to a small data burst, that is, burst $i$ contains the state ${\bf s}(\tau_i)$. Based on the received telemetry, the remote controller maintains a state estimate $\tilde{\bf s}(t)$. Clearly, as the UAV emits telemetry only at the time instants $\{\tau_i\}_{i{=}1,2,3,\ldots}$, even in an idealized case where $\Delta_i{=}0$ and $\omega_i{=}1$, $\forall i$, there is a mismatch between the actual and estimated state of the UAV due to the evolution of the state in between telemetry emission. The network transformation might increase the estimation error by delaying telemetry delivery and erasing some updates due to packet failure. Intuition may suggest that more frequent updates would decrease the estimation error. However, frequent updates may increase network congestion and buffer overload at the source.

\vspace{2mm}
\noindent
{\bf Computing Task Offloading:} As discussed earlier, the infrastructure can provide a deeper degree of assistance by taking over the execution of complex processing tasks.
For instance, a UAV may capture pictures in pre-set locations to detect events of interest (\emph{e.g.}, a car accident or suspicious human activities). In case of a positive detection, the UAV may be programmed to autonomously alter his course and acquire more information in that specific area. In a more extreme example of autonomy, the UAV may be closely pursuing a specific object based on sensor feed. In this case, as the UAV uses the outcome of processing to control fine-grain motion, the frequency of offloading is much higher compared to the previous example, and the requirements on delay and delay variations much more stringent. Clearly, the emission of processing tasks corresponds to the emission of a burst. The level of autonomy and the nature of the task (\emph{e.g.}, object detection on a video stream) determines the burst size distribution, as well as the delay and reliability requirements on the task delivery. In a typical use-case scenario, the on-board camera captures video with 720p HD resolution for video streaming, that requires a throughput of at least $1.8$~Mbps. Depending on the computation requirement, one can vary the frame rate or resolution and hence modify the task size. Usually, for tasks like image classification or object detection an entire image frame is sent as a burst of data to the server. For example, if images of average frame size $150 KB$ are captured with $10$ frames per second, and one application packet size is about 1500 bytes (including all headers) then a burst of $(150000/1500)= 100$ packets will be sent over network in every $100$ milliseconds.

\vspace{2mm}
The above case-study scenarios generate a broad spectrum of characteristics of infrastructure assistance in terms of both generated traffic -- frequency and size of bursts -- and requirements. Our simulation study attempts to bridge the UAV and communication/networking domains to provide a comprehensive discussion on the feasibility of infrastructure assistance to UAV operations.

\section{Simulation Environment}
\label{sec:simulator}


We briefly summarize the structure and main characteristics of the FlyNetSim simulator we presented in ~\cite{baidya2018flynetsim, flynetsimgit}, as well as the main general settings, used to obtain the results presented herein.

\subsection{FlyNetSim Simulator}

The objectives of FlyNetSim are: (\emph{i}) accurately model UAV operations and dynamics using a software-in-the-loop approach, where the data structures and control pipeline of UAV software are fully preserved; (\emph{ii}) accurately model a multi-scale multi-technology IoT communication environment and its interactions with the UAVs; (\emph{iii}) establish a one-to-one correspondence between UAVs and wireless nodes in NS-3, where the UAVs implement a full network stack supporting multiple network interfaces; (\emph{iv}) preserve individual data paths from and to UAV sensors and controllers.

To this aim, FlynetSim takes as starting point two open source simulators -- Network Simulator (NS-3) and ArduPilot -- and build a fully open source simulation environment for academic research. NS-3~\cite{henderson2008network} is an extremely popular tool, with a wide community contributing to extend its capabilities. ArduPilot~\cite{team2016ardupilot} is a widely used software and hardware-in-the-loop simulator, capable of modeling a broad range of unmanned vehicles characteristics in terms of navigation, control and mission planning.

FlyNetSim includes a middleware layer to interconnect the two simulators, providing temporal synchronization between network and UAV operations, and a publish and subscribe based framework~\cite{eugster2003many} to create end-to-end data-paths across the simulators. The middleware architecture we developed is lightweight, and enables FlyNetSim to simulate a large number of UAVs and support a wide range of IoT infrastructures and applications.

We refer the interested reader to~\cite{baidya2018flynetsim} 
for a in-depth description of the simulator and case-studies illustrating its capabilities.

\subsection{Urban Environment}



Based on the discussion presented earlier in this paper, it is clear that infrastructure assistance is a key enabler of autonomous UAV technologies. However, the UAVs and the infrastructure are necessarily connected through volatile wireless links. Especially in urban environments, parameters such as network load, signal propagation and mobility may degrade the ability of these links to reliably support assistance. In the following, we briefly describe the most salient aspects we explore in our simulations.

\subsubsection*{Propagation}
 In urban environments, high buildings often induce Non-Line of Sight (NLoS) signal propagation, where penetration loss and reflection through on building walls can cause highly varying attenuation. The dimension and location of the buildings, the thickness and materials of the walls also contribute to determine the overall amplitude of the received signals. The reliability of infrastructure assistance can be greatly impaired by these characteristics. 

To model these properties, we adopt a pathloss model 
for urban environments built as the combination of the standard ITU R1411 pathloss model~\cite{itu2015p} and other variable components function of salient parameters. Specifically, the path loss is defined as :

\vspace{-2mm}
\begin{equation}
\vspace{-2mm}
L = L_{\it b} + L_{\it ew} + G_{\it h} ,
\end{equation}
where $L_{\it ew}$ is the loss through the external walls and $G_{\it h}$ is the gain due to the altitude of the device. The loss $L_{\it b}$ is the basic pathloss function
\begin{equation}
L_{\it b}=\left\{
  \begin{array}{@{}ll@{}}
    L_{\it los}, & \text{if}\ Line~of~sight \\
    L_{\it nlos}, & \text{otherwise}
  \end{array}\right.
\end{equation}
In line of sight, the pathloss term is defined as:
\begin{equation}
L_{\it los} = \bigm|20log\bigm(\frac{\lambda^2}{8\pi h_{bs}h_{uav}})\bigm| + ~C  ,
\end{equation}
where $h_{bs}$ and $h_{uav}$ are the altitudes of the base station and UAV respectively,
$\lambda$ is the wavelength and $C$ is a variable whose upper and lower bound are a function of distance, wavelength and altitude of the devices.
In LoS, the pathloss term is defined as:

\vspace{-2mm}
\begin{equation}
\vspace{-2mm}
L_{\it nlos} = L_{\it fs} + L_{\it df},
\end{equation}
where $L_{\it fs}$ is the free space pathloss and $L_{\it df}$ is the total pathloss due to  diffraction, including that generated by the roof top and from the rows of the buildings. The value of $L_{\it df}$ depends on the frequency of the signal, distance of transmitter and receiver, and altitude of the devices. Note that urban pathloss not only restricts the communication range, but also has large impact on resource allocation and modulation schemes in the communication. In our feasibility study, we consider non-LoS pathloss model.


 \subsubsection*{Mobility}
 The maximum speed of UAVs can be of the order of 100 mph and the mobility patterns depend on several factors including the mission objective and environment. Unlike MANET, the UAVs move in the network in a 3D space leading to an increased range of dynamics, and the high speed leads to a faster change in terms of channel gain. In infrastructure-based UAV network, UAV may go out of coverage of the current base station or access point. However, most of the infrastructures  provide handover mechanisms for continued connectivity.

 \subsubsection*{Exogenous Traffic}
 In case of infrastructure-assisted UAV networks, the mission-based application data stream is affected by the contention of data streams generated by users using the same access point or base station. 
 The infrastructure can provide flow control and congestion control, and attempt to achieve fairness in resource allocation among coexisting applications and communicating devices.

 
Note that interference can be generated by coexisting wireless networks or the presence of local Device-to-Device (D2D) communications using the same frequency band. The infrastructure can implement a centralized or distributed coordination in terms of transmission scheduling and power control to avoid or mitigate the impact of interference. Especially when heterogeneous network technologies share same spectrum causing interference, the different technologies can cooperate through the core backbone network to improve overall performance.

\subsubsection*{Network Technology}

Technologies and protocols are often designed under certain assumptions and tailored to specific scenarios. The technologies in urban environments have different specifications and approaches to data transmission and channel sharing, e.g., WiFi employs a distributed algorithm for contention based resource allocation, whereas in LTE resource allocation is controlled using a centralized policy by the base station. The technologies also differ in terms of retransmission strategy, error correction, interference control, security and many other aspects which advocates to choose appropriate technology and protocol in different application scenarios and different network conditions.

\section{Results and Discussion}
\label{sec:results}

\begin{figure}[t]
\centering
\vspace{-2mm} 
\includegraphics[width=0.8\linewidth]{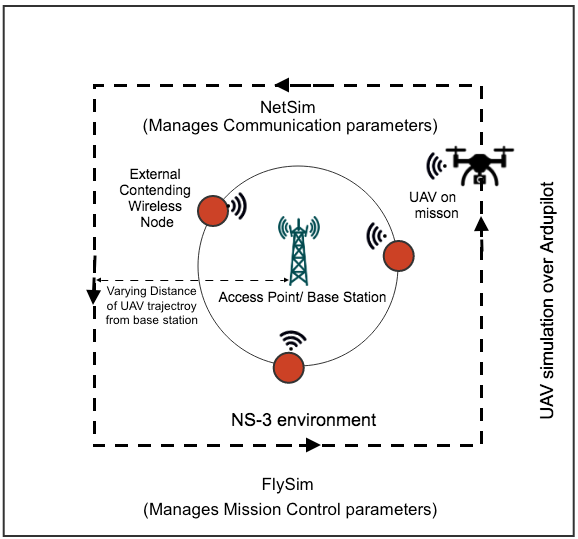}
\caption{Simulation setup: the UAV follows a predefined trajectory, whereas external nodes are fixed and disposed in a circle centered on the AP or eNodeB.}%
  \label{fig:simulation_setup}%
\vspace{-6mm}  
\end{figure}

\subsection{Simulation Setup}
We simulate the environment and applications described earlier using the integrated UAV-network simulator FlyNetSim. The simulated UAV performs a predefined mission, which corresponds to a navigation plan in an urban environment. The wireless environment is created using ns-3, the network simulator component of FlyNetSim. The mobility of the vehicle is updated in the network simulator in real time and network parameters are varied to evaluate the performance in a range of scenarios. A realistic simulation of motion and control and control of the UAV is obtained using ArduPilot. The simulated UAV uses simulated sensors to provide telemetry information to the GCS over a simulated network. 
In the second application, the UAV generates tasks/packet bursts with different statistics and uses simulated WiFi and LTE networks to transport them to an edge server. 

Ground WiFi/LTE nodes are added with uniform disc position allocation close to the GCS node. These nodes produce data streams directed to the same access point or eNodeB. The number of external nodes, the datarate and the packet size of the ground users are tunable parameters.
The pathloss and shadowing parameters are set based on the ItuR1411 Propagation Loss Model.  Motion, shadowing and exogenous traffic all affect the transformation induced by the network on the packet stream from the UAV. The parameters used in the simulation are summarized in table~\ref{table:params}. The UAV moves in a rectangular trajectory around the AP or base station as shown in Figure~\ref{fig:simulation_setup}.


\begin{table}[!t]
\begin{tabular}{p{4cm}p{4cm}}
\small

     \begin{tabular}{|l|l|}
        \hline
        Parameters & Value\\
        \hline
        UAV Mobility & Constant Speed\\
        WiFi Standard & IEEE 802.11a\\
        WiFi Bandwidth & 20 MHz\\
        Propagation loss Model & ItuR1411 Propagation Loss\\
        Propagation Delay Model & Constant Speed\\
        LTE EARFCN & 18000\\
        LTE Bandwidth & 20 MHz\\
        LTE RLC Mode & Acknowledgement (RLC AM)\\
        LTE downlink MAC Scheduler & Proportional Fair\\
        LTE uplink MAC Scheduler & Round Robin\\
        TCP Congestion Control & New Reno\\
        
        \hline
        \end{tabular}     
\end{tabular}
\caption{Experiment Parameters used in the simulations.}
\vspace{2mm}
\label{table:params}
\vspace{-4mm}
\end{table}

\subsection{Remote Navigation Assistance}

\begin{figure}[t]
\includegraphics[width=\linewidth]{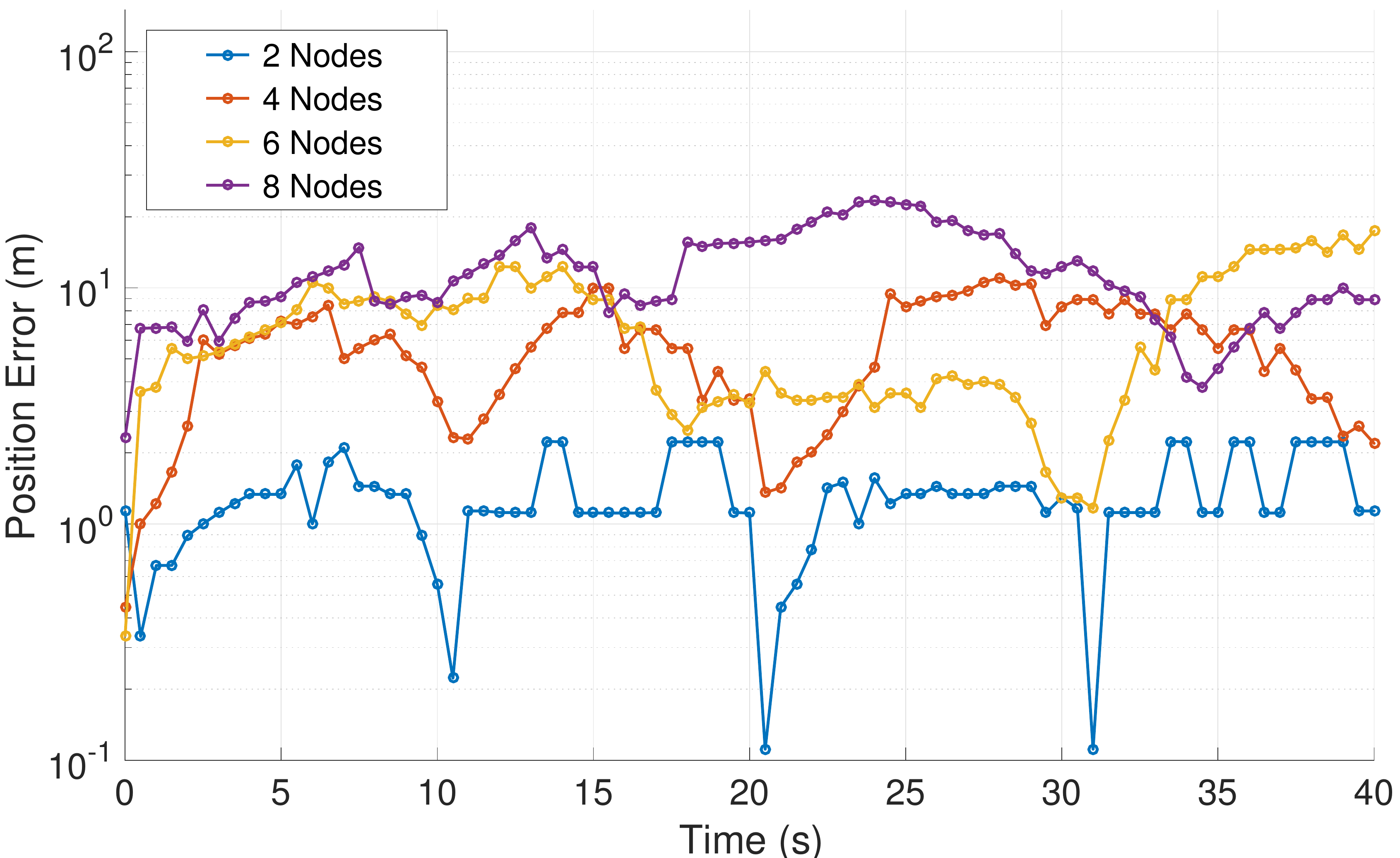}
\caption{Temporal evolution of position error based on the telemetry emitted by the UAV over WiFi with TCP in an urban environment with buildings. Each external node transmits traffic with rate 6 Mbps. The UAV navigates on a rectangular trajectory where the access point is at the center.}%
  \label{fig:wifi_tcp_pos_er_temporal}%
\vspace{-4mm}   
\end{figure}

Figure~\ref{fig:wifi_tcp_pos_er_temporal} shows an example of temporal evolution of error in position estimation while the UAV performs a mission over the predefined trajectory. In this experiment, communication is over WiFi, and the different lines correspond to a different number of nodes contending the shared channel resource with the UAV. 
The contending nodes are placed with uniform disc position allocation in close distance from the AP/base station and transmits periodic bursts of packets of size $800$ bytes with uniform periodicity based on the traffic rate; e.g., for traffic rate of 6 Mbps, the packets are transmitted at interval of $(800*8)/6 = 1067 {\mu}s $.
It can be seen how not only the mean of estimation error grows as the number of contending nodes increases, but also the variance. The line corresponding to $2$ nodes has few sections where the error doubles, probably caused by TCP timeouts, but otherwise shows a considerable stability. Note that the error has periodic low spikes, corresponding to periods of time where the UAV is stationary. A higher number of nodes contending the wireless resource further increase average and variations of delay, with extended sections of the trace where the error significantly increases with respect to the minimum. This latter effect maps to a general inability of a remote node to maintain a tight control over UAV trajectories.

We measure the average performance of telemetry transmission for navigation and computation task offloading in various scenarios. For the external traffic, we define two extreme regimes - \emph{no load} conditions, where no wireless node other than the UAV is present, and \emph{high load} conditions, where 8 contending nodes are present generating $6$~Mbps in the application layer. We also define \emph{Low Distance} and \emph{High Distance} scenarios, where the distance between the UAV and the access point or base station is $10$ and $40$~m, respectively.

\begin{figure}[t]
\includegraphics[width=\linewidth]{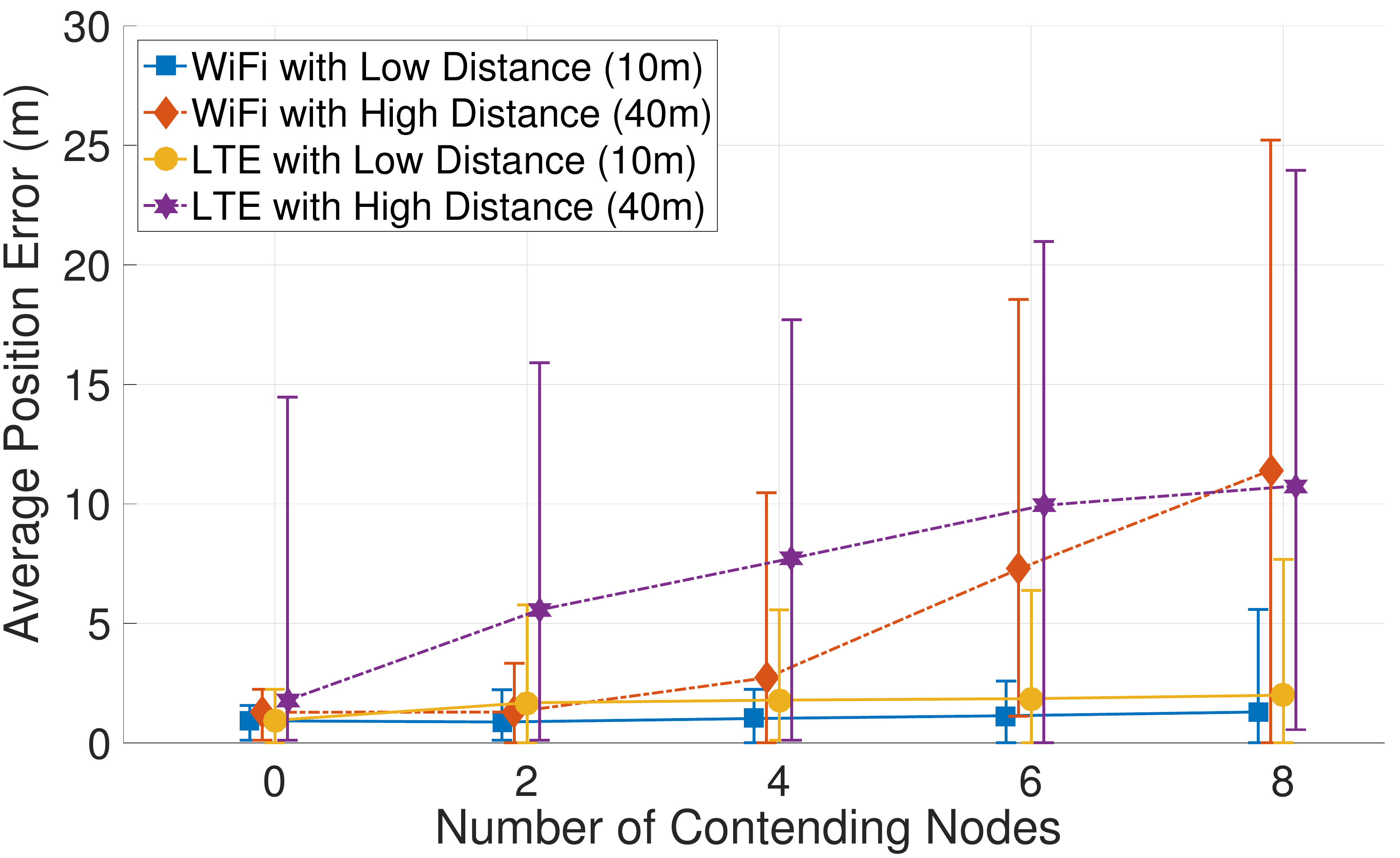}
\caption{Average position error and variation range for different communication technologies and protocols with respect to varying number of contending nodes. Each node is transmitting 6 Mbps data traffic.}%
  \label{fig:avg_pos_err_vs_nodes}%
\vspace{-4mm} 
\end{figure}

Figure~\ref{fig:avg_pos_err_vs_nodes} compares the average position error \emph{High} and \emph{Low Distance} conditions over WiFi and LTE networks as a function of the number of ground nodes. It can be seen how both networks maintain a low error as the number of nodes grow, with WiFi being a slightly better option. At $40$~m, WiFi has an manifestly lower error compared to LTE in mild contention environments, but presents a sharp degradation as the number of ground nodes grows. This is due to the inefficiency of DCF and random access in high-load conditions compared to the LTE more controlled MAC.

\begin{figure}[t]
\includegraphics[width=\linewidth]{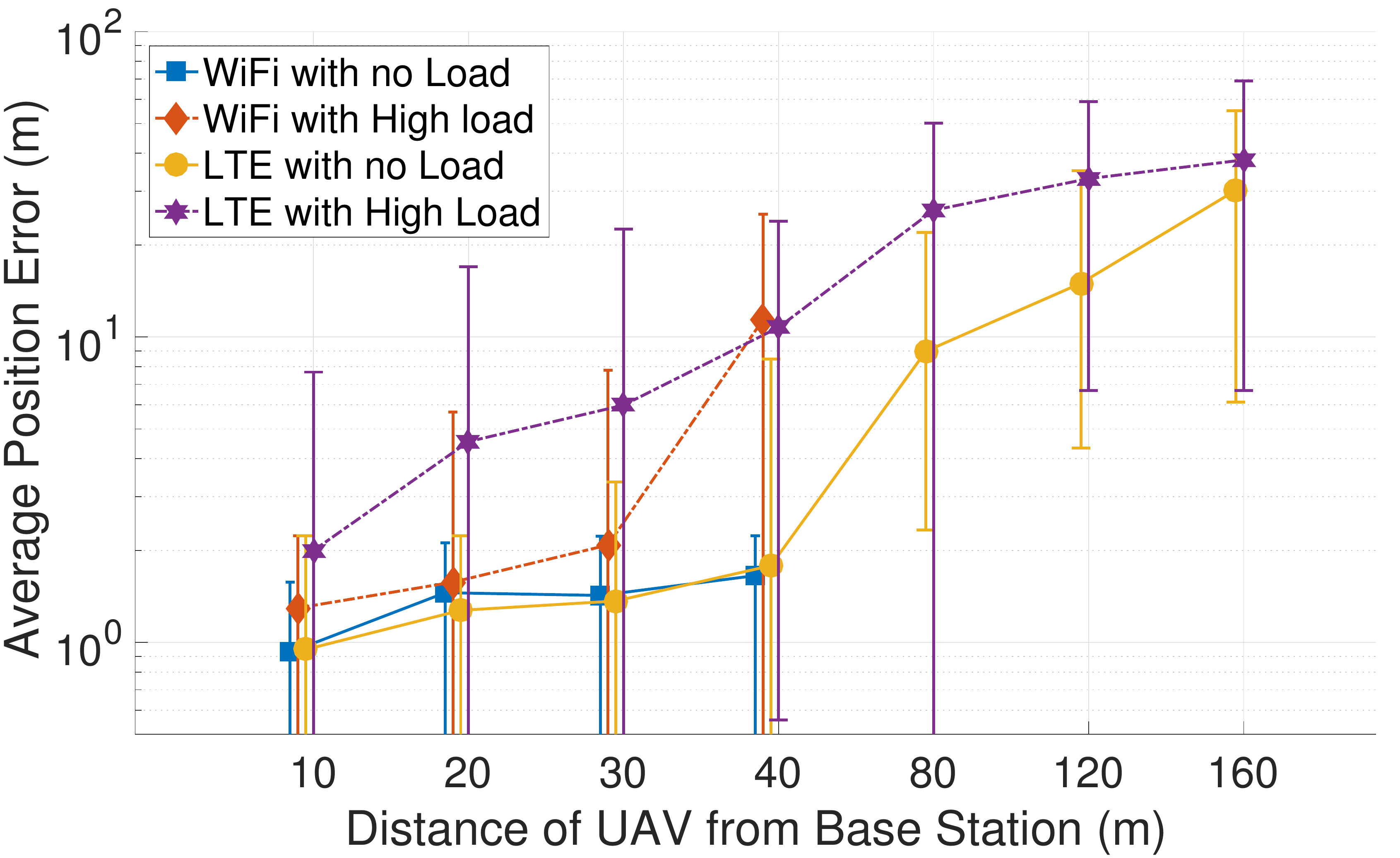}
\caption{Average position error and variance as a function of distance between the UAV and the base station for different communication technologies. The measurements are taken in no load and high load (8 nodes 6 Mbps each) conditions.}%
  \label{fig:avg_pos_err_vs_dist}%
\vspace{-4mm}  
\end{figure}

Figure~\ref{fig:avg_pos_err_vs_dist} shows the same metric over WiFi and LTE in \emph{High} and \emph{No Load} conditions as a function of distance. Note that $40$~m is the disconnection limit for WiFi, whereas LTE has a much extended range, although at the price of a large position error. The effect of distance and load is apparent in both technologies. However, it can be seen how in the absence of traffic from other nodes the two options are essentially equivalent until disconnection. Conversely, in \emph{High Load} conditions the difference is marked: WiFi has a much lower error at moderate distance, and then sharply degrades as the UAV approaches the maximum range.

\begin{figure}[t]
\includegraphics[width=\linewidth]{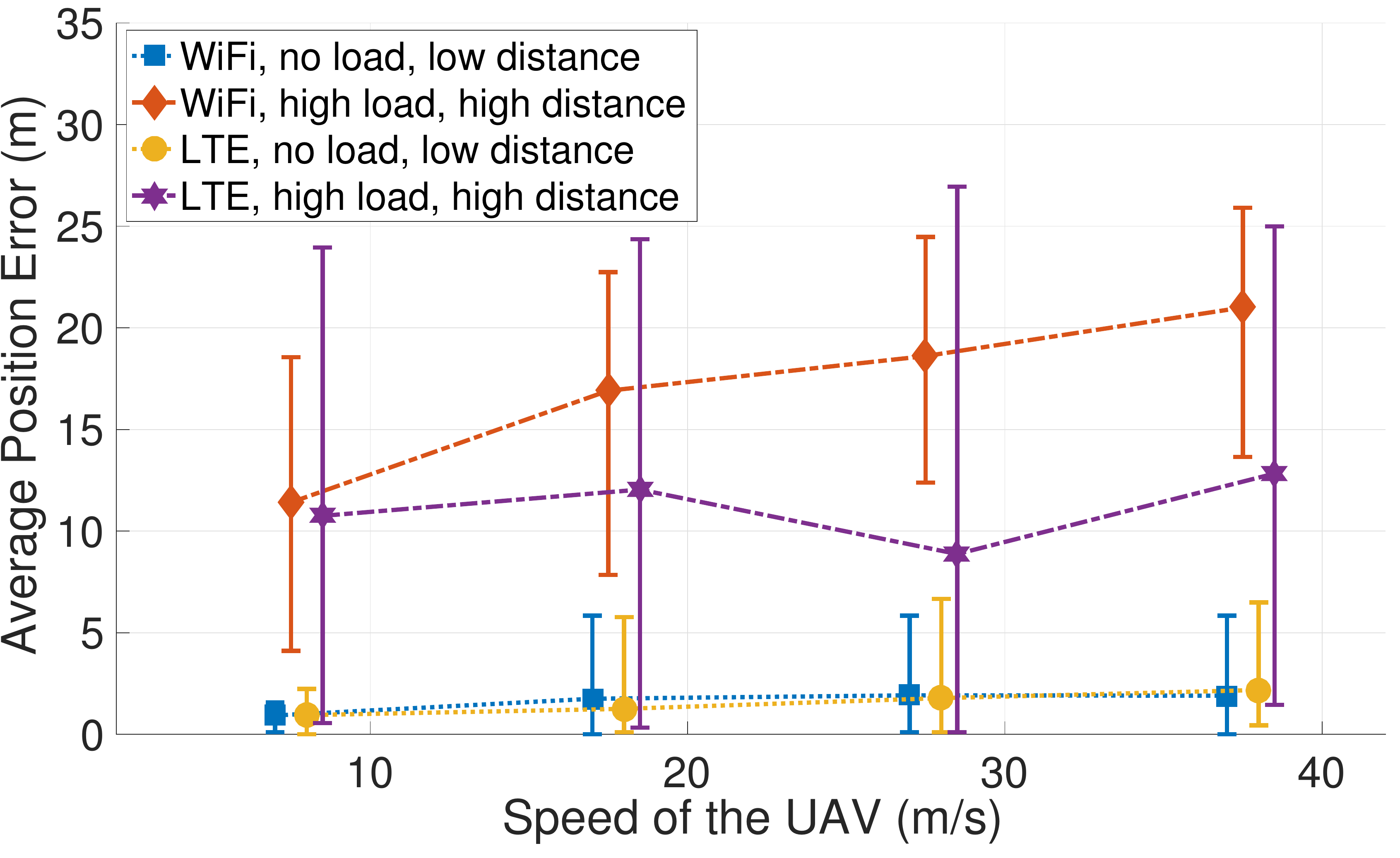}
\caption{Variation of Average position error with the speed of the UAV in different network load conditions over LTE and WiFi.}%
  \label{fig:avg_pos_err_vs_uav_speed}%
\end{figure}

\begin{figure}[t]
\includegraphics[width=\linewidth]{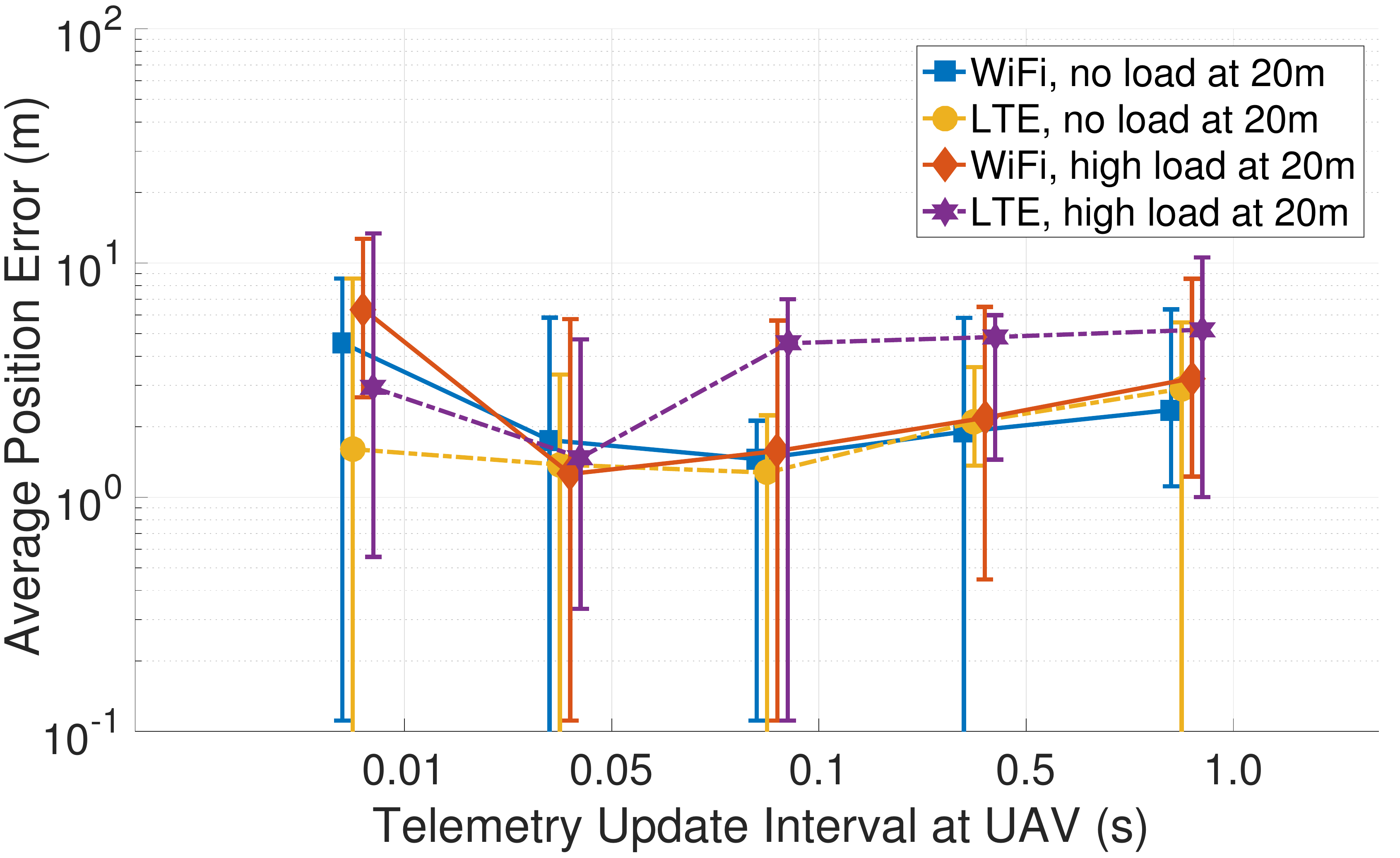}
\caption{
Variation of Average position error with different update frequency at the UAV for telemetry over WiFi and LTE. The measurements are taken in two regimes: no traffic load and UAV is at low distance (10m), and high traffic load (6 nodes, 6 Mbps each) when UAV is at high distance (40m).}%
  \label{fig:avg_pos_err_vs_update_freq}%
\vspace{-4mm}   
\end{figure}

We also measure the impact of speed on the position error. The plot is in \emph{High Distance} regime. Intuitively, a higher speed of the vehicle would result in a larger error simply due to the fact that the UAV would have moved farther since a packet containing the last update was received. Figure~\ref{fig:avg_pos_err_vs_uav_speed} shows the effect of increasing speed of UAV in \emph{High} and \emph{No Load} conditions. As expected, the absolute position error increases with the vehicle speed. The impact of load is also manifest: the higher inter-packet delay maps to a faster error increase. 



We also measure how the frequency of updates from the the UAV affects the absolute value of the position estimation error.  Figure~\ref{fig:avg_pos_err_vs_update_freq} shows the variation at distance (20m). Intuitively, a low update frequency generates a small network load, but also allows the UAV to travel farther in between updates. Conversely, frequent updates means a smaller error in idealized conditions, but impose a larger load to the network. This trend is shown in the plot, where the error has a minimum which depends on the technology and external load. Note that the effect of the additional load introduced by frequent updates is more pronounced in LTE \emph{High Load}, due to the smaller maximum throughput of the network in those conditions.


\begin{figure}[t]
\includegraphics[width=\linewidth]{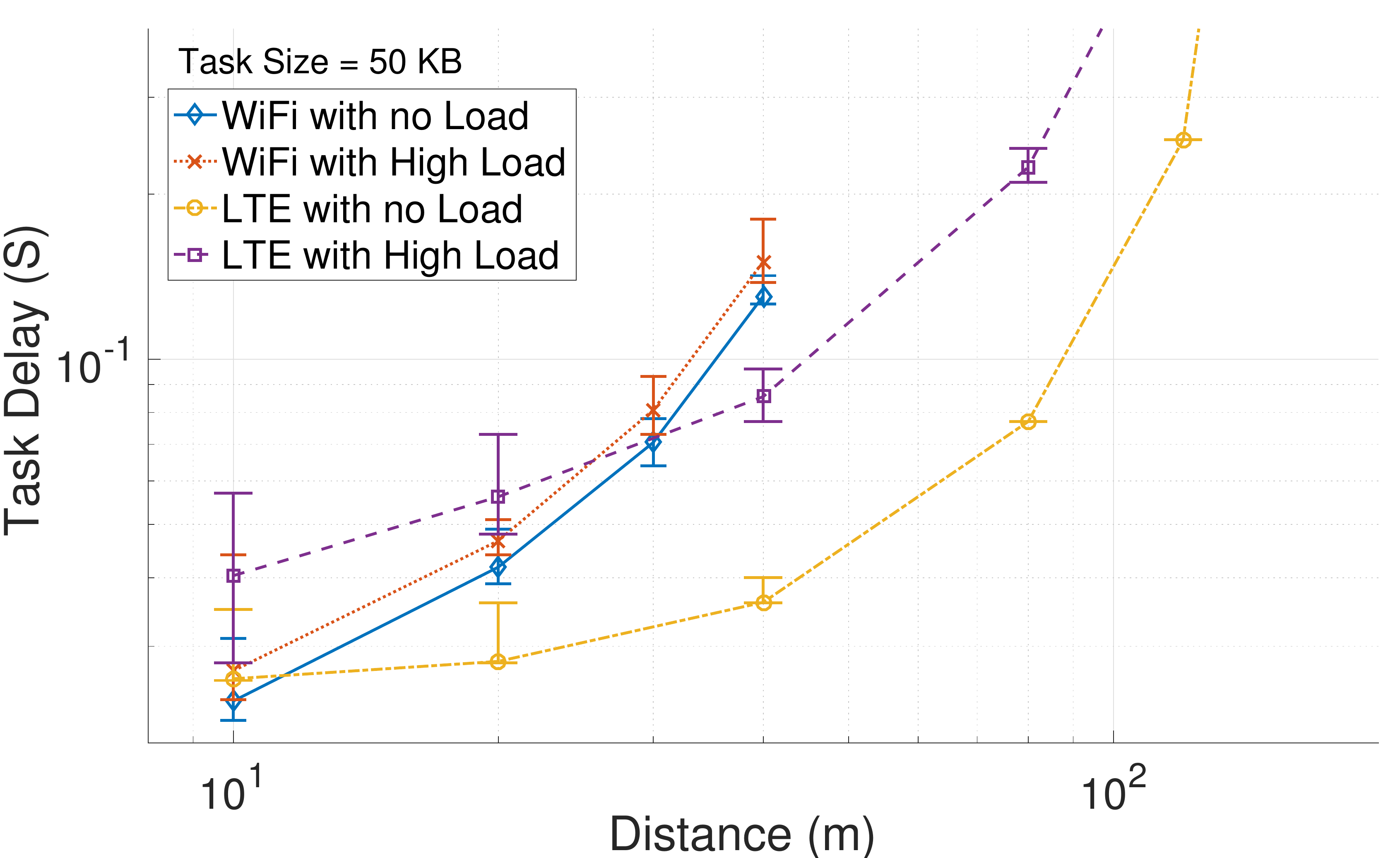}
\caption{Task delay over WiFi and LTE with varying distance of the UAV from the Base Station for task of burst size 50 KB every second. The measurement are taken in two different external traffic regimes of no load and high load.}%
  \label{fig:task_vs_dist_50}%
\end{figure}

\begin{figure}[t]
\includegraphics[width=\linewidth]{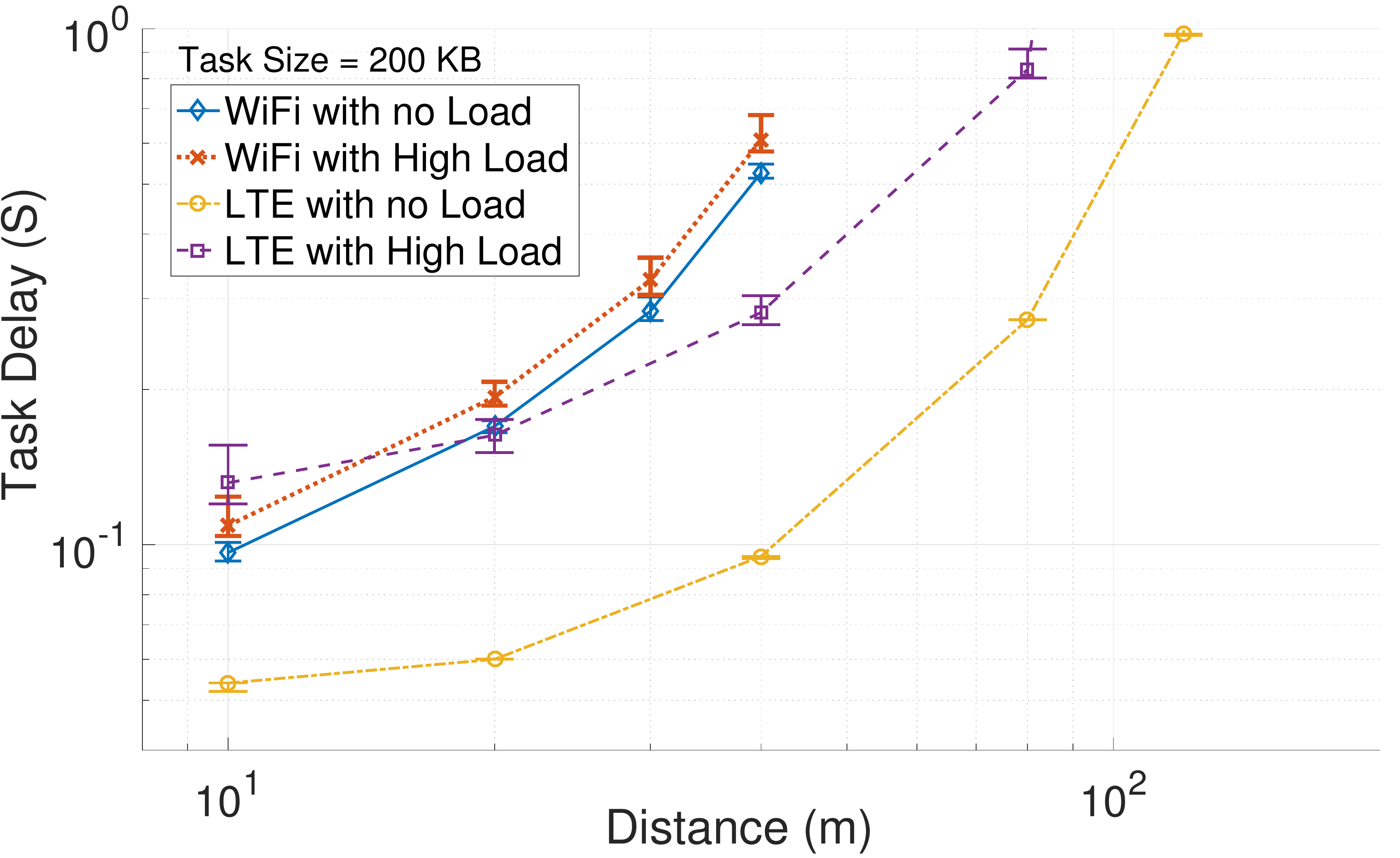}
\caption{Task delay over WiFi and LTE with varying distance of the UAV from the Base Station for task of burst size 200 KB every second. The measurement are taken in two different external traffic regimes of no load and high load}%
  \label{fig:task_vs_dist_200}%
\vspace{-4mm}   
\end{figure}


\subsection{Computing Task Offloading}

We now focus on evaluation of the performance when the UAV is transmitting burst of packets with certain characteristics.

Figure~\ref{fig:task_vs_dist_50} shows the variation of task delay as a function of distance in a scenario where the tasks correspond to small $50$~KB data are transmitted from the UAV every second. Interestingly, it can be observed that in this case load conditions have a small impact on WiFi, whereas LTE suffers a larger number of nodes using the same channel. LTE clearly outperforms WiFi in \emph{No Load} conditions unless the UAVs is very close to the access point. Conversely, WiFi has a smaller error compared to LTE in \emph{High Load} conditions up to $30$~m distance, where the smaller range of WiFi penalizes this choice.
These results are strongly influenced by the small size of the task, which makes the corresponding packets go through transmissions in the WiFi MAC. In LTE, the round robin allocation of resources increases the delay in the presence of other users. 

Figure~\ref{fig:task_vs_dist_200} shows the same plot where tasks are of size $200$~KB. It can be observed a general shift of all the delays, with WiFi being the most penalized by the size increase.
 One of the reason for LTE incurring a less perceivable degradation with respect to WiFi is the RLC buffer, that reduces the retransmission at the TCP layer~\cite{kumar2018dynamic} compared to WiFi where the large burst size can cause more back-offs. However, when the network is congested, resource allocation still penalizes LTE in uplink as round robin is used.

\begin{figure}[t]
\includegraphics[width=\linewidth]{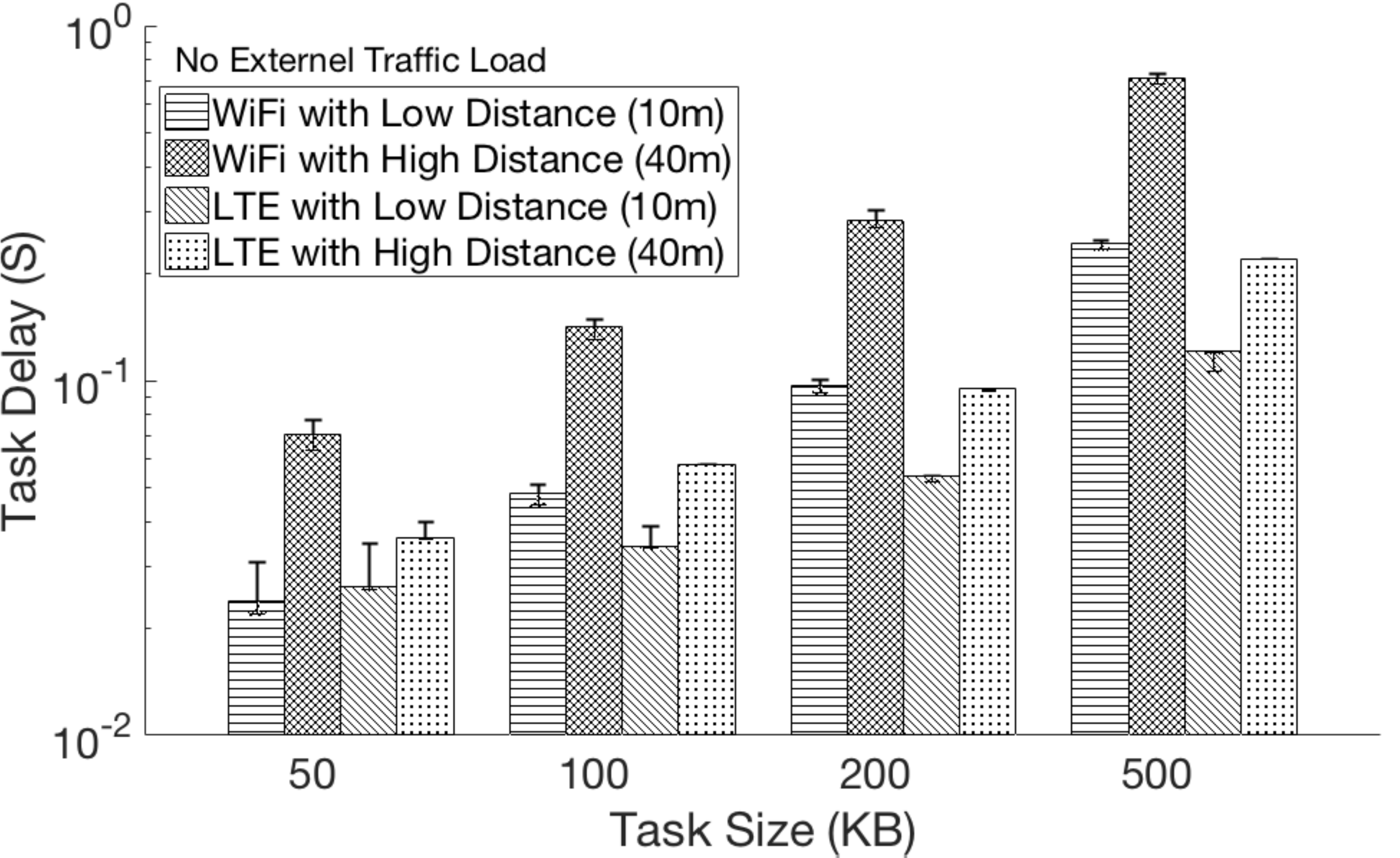}
\caption{Task delay over WiFi and LTE with varying task sizes for low and high distance of the UAV from the Base Station. The measurement are taken in in absence of any external traffic load.}%
  \label{fig:task_vs_size_no_load}%
\end{figure}

\begin{figure}[t]
\includegraphics[width=\linewidth]{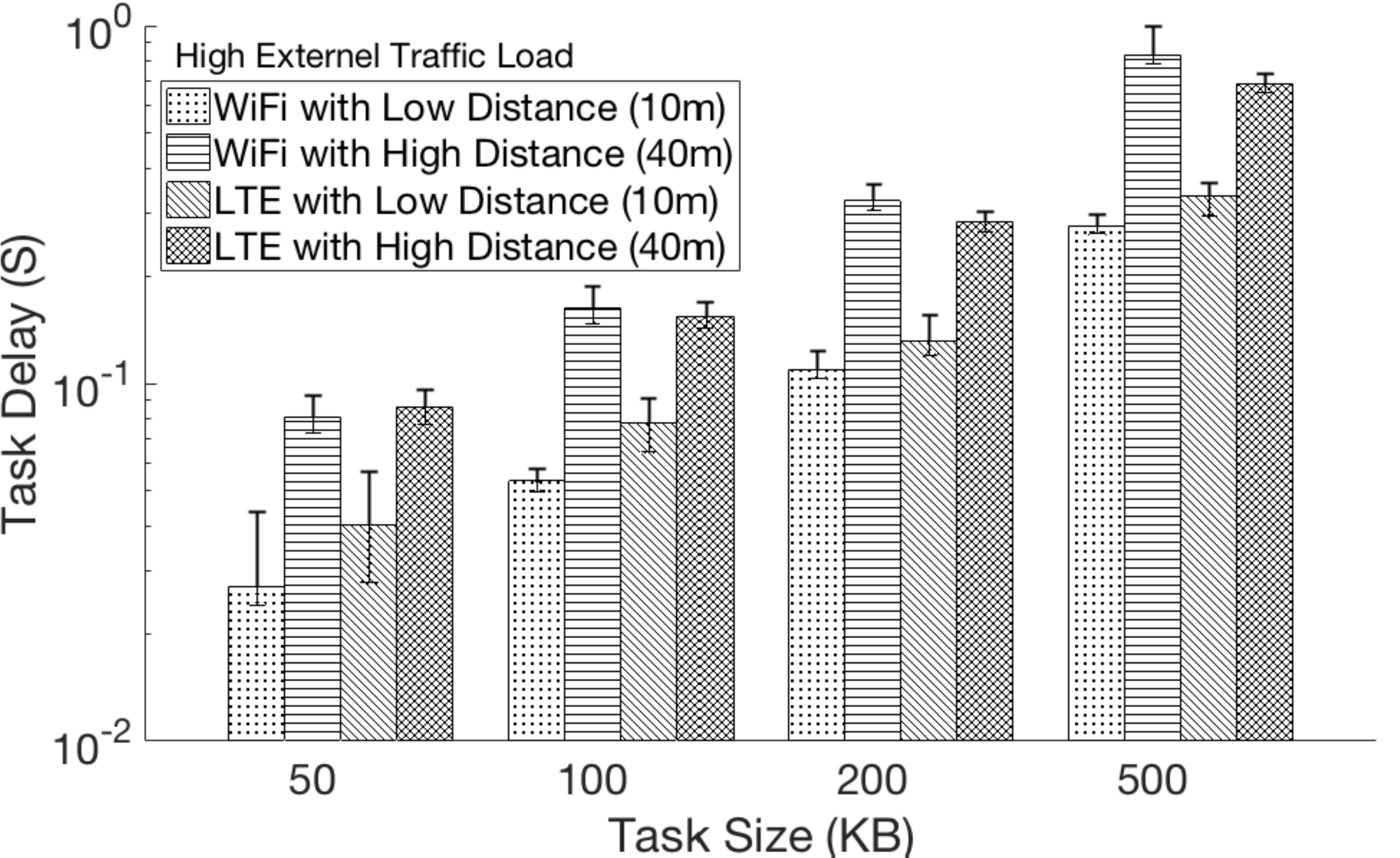}
\caption{Task delay over WiFi and LTE with varying task sizes for low and high distance of the UAV from the Base Station. The measurement are taken in in presence of external traffic load of 3 nodes transmitting traffic of 1 Mbps each.}%
  \label{fig:task_vs_size_high_load}%
\vspace{-4mm}   
\end{figure}

We give a more clear view of this effect in Figure~\ref{fig:task_vs_size_no_load}, where we show delay as a function of task size in \emph{No Load} conditions. WiFi suffers a larger delay at high distance, but provides better performance compared to LTE in the short range. It can be seen how WiFi has a steeper delay increase as the task size increases.

Finally, Figure~\ref{fig:task_vs_size_high_load} shows that in presence of external traffic in the network, WiFi has a smaller delay compared to LTE in closer range for all task sizes. This is due to the fact that the task is transmitted over LTE uplink which schedules in round robin all the UE nodes data. Also, it adds additional delays every time it needs to seek transmission opportunity from the eNodeB for a new chunk of data from any given UE. However, in longer distance, WiFi deteriorates fast and slightly worsen the performance compared to LTE.

\subsection{Discussion}
Based on the results from two use case scenarios of infrastructure-assisted UAV, we can see that the choice of network depends on the network conditions, distance from the access point or base station, as well as the class of application that UAV is serving. In remote navigation assistance, the telemetry data are more suitable to be transmitted over WiFi when the UAV is close to the AP. In applications such as task offloading, the size of the task has a great impact on the network to be used. When large data bursts are to be transmitted efficiently to the edge with low latency, the LTE network provides best performance in low-load conditions. However, if the network is congested, connecting the UAV over WiFi is still advantageous at short distance from the AP, whereas LTE is the best option. We conclude that in order to achieve efficient network infrastructure-assistance, the UAV should be multi-homed, that is, it should have both WiFi and LTE interfaces and use a context and application aware policy to determine which network should be used.

\section{Conclusions}
\label{sec:conclusion}

The main objective of this paper is to provide a comprehensive evaluation of communication strategies for infrastructure assistance to the operations of autonomous UAVs. Based on detailed UAV-network simulations, we focus our attention on remote navigation assistance and offloading of processing tasks to edge servers. Our results, obtained using the recently proposed UAV-network simulator FlyNetSim, indicate the need for the UAVs to be equipped with multiple network interface and be able to switch from one to another during missions.

\bibliographystyle{IEEEtran}
\bibliography{infra_assist}

\end{document}